\documentclass{cmmse2008}
\usepackage{amssymb}

\usepackage{graphics}
\usepackage{fancybox,here,epsf}

\usepackage{amsfonts}
\usepackage{amssymb}
\usepackage{amsmath,latexsym}

\begin{document}

\pagestyle{empty}




\newcommand\epsfig[4]{\begin{figure}[#2]
  \centerline{
    \epsfysize=#1
    \epsfbox{#3.eps}
    }
\vspace{0.2cm}
\caption{\em #4 \label{fig:#3}}  
\end{figure}
}


\newcommand\epsfigtwo[6]{\begin{figure}[#3]
  \centerline{
   a)  \epsfysize=#1
       \epsfbox{#4.eps} $ \quad \quad \quad $
   b)  \epsfysize=#2
       \epsfbox{#5.eps}
       }
\caption{{\em #6} \label{fig:#4}}  
\end{figure}
}


\newcommand\epsfigthreevar[8]{\begin{figure}[#4]
  \centerline{
   a)  \epsfysize=#1
       \epsfbox{#5.eps} $ \quad  $	}
\vspace{10mm}
\centerline{
   b)  \epsfysize=#2
       \epsfbox{#6.eps} $ \quad  $	
   c)  \epsfysize=#3
       \epsfbox{#7.eps} $ \quad  $	}
\caption{#8 \label{fig:#5}}  
\end{figure}
}


\newcommand\epsfigthree[8]{\begin{figure}[#4]
  \centerline{
   a)  \epsfysize=#1
       \epsfbox{#5.eps} $ \quad  $	
   b)  \epsfysize=#2
       \epsfbox{#6.eps} $ \quad  $	
   c)  \epsfysize=#3
       \epsfbox{#7.eps} $ \quad  $	}
\caption{{\em #8} \label{fig:#5}}  
\end{figure}
}


\newcommand\epsfigfour[8]{\begin{figure}[#3]
  \centerline{
   a)  \epsfysize=#1
       \epsfbox{#4.eps} $ \quad  $      
   b)  \epsfysize=#1
       \epsfbox{#5.eps} $ \quad  $      }
\centerline{
   c)  \epsfysize=#2
       \epsfbox{#6.eps} $ \quad  $      
   d)  \epsfysize=#2
       \epsfbox{#7.eps} $ \quad  $      }
\caption{\em #8 \label{fig:#4}}  
\end{figure}
}



\newcommand{\Eqref}[1]{(\ref{#1})}


\markboth
  {Template for the CMMSE 2008 proceedings} 
  {Regino Criado, J. Vigo Aguiar}


\title{Torus bifurcations, isolas and chaotic attractors\\
  in a simple dengue model with ADE \\ and temporary cross immunity}



\author{Ma\'ira Aguiar}{maira@igc.gulbenkian.pt}{1}

\author{Nico Stollenwerk}{nico@ptmat.fc.ul.pt}{1}

\author{Bob W. Kooi}{kooi@falw.vu.nl}{2}


\affiliation{1}{Centro de Matem\'atica e Aplica\c{c}\~oes Fundamentais}
  {Universidade de Lisboa, Portugal}

\affiliation{2}{Faculty of Earth and Life Sciences, 
Department of Theoretical Biology}
{Vrije Universiteit Amsterdam, Nederland}


\begin{abstract}
We analyse an epidemiological model of competing strains of pathogens
and hence differences in transmission for first versus secondary
infection due to interaction of the strains with previously
aquired immunities, as has been
described for dengue fever 
(in dengue known as antibody dependent enhancement, ADE). 
Such models show a rich variety of dynamics 
through bifurcations up to deterministic chaos. Including temporary
cross-immunity even enlarges the parameter range of such chaotic attractors,
and also gives rise to various coexisting attractors, which are difficult
to identify by standard numerical bifurcation programs using continuation
methods. A combination of techniques, including classical bifurcation plots
and Lyapunov exponent spectra has to be applied in comparison to
get further insight into such dynamical structures.
Here we present for the first time multi-parameter studies
in a range of biologically plausible values for dengue. 
The multi-strain interaction with the immune system is expected to also
have implications for the epidemiology of other diseases.

\keywords numerical bifurcation analysis, Lyapunov exponents,
$\mathbb{Z}_2$  
symmetry, coexisting attractors, 
antibody dependent enhancement (ADE)
\end{abstract}

%
%

\section{Introduction}

Epidemic models are classically phrased in ordinary differential
equation (ODE) systems for the host population divided in classes of
susceptible individuals and infected ones (SIS system), 
or in addition, a class of recovered individuals due to immunity after
an infection to the respective pathogen (SIR epidemics).
The infection term includes a product of two variables, hence
a non-linearity which in extended systems can cause
complicated dynamics. Though these simple SIS and SIR models
only show fixed points as equilibrium solutions, they already
show non-trivial equilibria arising from bifurcations, and in
stochastic versions of the system critical fluctuations at the
threshold. 
  Further refinements of the SIR model in terms of external 
forcing or distinction of infections with different strains
of a pathogen, hence classes of infected with one or another
strain recovered from one or another strain, infected with
more than one strain etc., can induce more complicated dynamical
attractors including
equilibria, limit cycles, tori and chaotic attractors.
  
  Classical examples of chaos in epidemiological models are childhood
diseases with extremely high infection rates, so that a moderate
seasonal forcing can generate Feigenbaum sequences of period
doubling bifurcations into chaos. 
The success in analysing childhood
diseases in terms of modelling and data comparison lies in the fact that
they are just childhood diseases with such high infectivity. Otherwise
host populations cannot sustain the respective pathogens.
In other infectious diseases much lower forces of infection
have to be considered leading to further conceptual problems with
noise affecting the system more than the deterministic part, leading 
even to critical fluctuations with power law behaviour, when
considering evolutionary processes of harmless strains of pathogens
versus occasional accidents of pathogenic mutants 
\cite{meningsoc2003}.
Only explicitly stochastic models, of which the classical ODE models
are mean field versions, can capture the fluctuations 
observed in time series data
\cite{pnasscript}.

  More recently it has been demonstrated that the interaction of
various strains on the infection of the host with eventual
cross-immunities or other interactions between host immune system and
multiple strains can generate complicated dynamic attractors.
A prime example is dengue fever. A first infection
is often mild or even asymptomatic and leads to
life long immunity against this strain. However, 
a subsequent infection with another strain of the virus
often causes
clinical complications
up to life threatening conditions and 
hospitalization, due to ADE.
More on the biology of dengue and its consequences for the detailed
epidemiological model structure can be found 
in Aguiar and Stollenwerk \cite{AgSt2007} including
literature on previous modelling attempts, see also
\cite{EduardoMassad2008}. On the biological evidence for ADE see e.g.
\cite{Halstead2003}.
Besides the difference in the force of infection between primary
and secondary infection, parametrized by a so called ADE 
parameter $\phi $, which has been
demonstrated to show chaotic attractors in a certain parameter region,
another effect, the temporary cross-immunity after a first
infection against all dengue virus strains, parametrized by
the temporary cross-immunity rate
$\alpha $, shows bifurcations up to chaotic attractors in a much wider
and biologically more realistic parameter region.
  The model presented in the Appendix has been described
in detail in \cite{AgSt2007} and has recently been analysed for a parameter
value of $\alpha = 2 \; year ^{-1} $ corresponding to on average half
a year of temporary cross immunity which is biologically plausible
\cite{DengueChaos2008}. For increasing ADE parameter $\phi $ first an
equilibrium which bifurcates via a Hopf bifurcation into a stable
limit cycle and then after further continuation the limit cycle
becomes unstable in a torus bifurcation. This torus bifurcation can be
located using numerical bifurcation software based on continuation
methods tracking known equilibria or limit cycles up to bifurcation
points \cite{autoreference}. The continuation techniques and the
theory behind it are described e.g. in
Kuznetsov \cite{Kuzn2004}.
Complementary methods like Lyapunov
exponent spectra can also characterize chaotic attractor
\cite{Ruelle89,ChaosinDynamicalSystems}, and led ultimately to the
detection of coexisting attractors to the main limit cycles and
tori originated from the analytically accessible fixed point
for small $\phi $.
Such coexisting structures are often missed in bifurcation analysis
of higher dimensional dynamical systems but are demonstrated
to be crucial at times in
understanding qualitatively the real world data, as for example
demonstrated previously in a childhood disease 
study \cite{drepperengstollenw}. In such a study first the understanding
of the deterministic system's attractor structure is needed, 
and then eventually the
interplay between attractors mediated by population noise
in the stochastic version of the system gives the full understanding of
the data.
  Here we present for the first time extended results 
of the bifurcation structure for various 
parameter values of the temporary cross immunity $\alpha $ in the
region of biological relevance and multi-parameter bifurcation
analysis. This reveals besides the torus bifurcation route to
chaos also the classical Feigenbaum period doubling sequence
and the origin of so called isola solutions. The symmetry of the
different strains leads to symmerty breaking bifurcations of
limit cycles, which are rarely described in the 
epidemiological
literature but well known in the biochemical literature, e.g
for coupled identical cells.
The interplay between different numerical procedures and basic
analytic insight in terms of symmetries help to understand the
attractor structure of multi-strain interactions in the present case of
dengue fever, and will contribute to the final understanding of
dengue epidemiology including the observed fluctuations in 
real world data.
In the literature the multi-strain interaction leading to
deterministic chaos via ADE has been described previously, e.g.
\cite{FergusonN.et.al1999,Billings2007}
but neglecting temporary cross immunity and hence getting
stuck in rather unbiological parameter regions, whereas more recently
the first considerations of temporary cross immunity in rather
complicated and up to now not in detail analysed models 
including all kinds of interations have appeared 
\cite{HelenRohani2006,YoshiNagaoKoelle2008}, in this case failing
to investigate closer the possible dynamical structures.

\vspace{-0.3cm}

\section{Dynamical system}

The multistrain model under investigation can be given as an
ODE system 
\begin{equation}
	\frac{d}{dt} \; \underline x = 
               \underline f (\underline x, \underline a)
	\label{dynamicsf}
\end{equation}
for the state vector of the epidemiological host classes
$
  \underline x :=(S,I_1, I_2, ... , R)^{tr} 
$
and besides other fixed parameters which are biologically undisputed
the parameter vector of varied parameters 
$\underline a = (\alpha, \phi)^{tr} $. For a detailed description
of the biological content of state variables and parameters see
\cite{AgSt2007}. The ODE equations and fixed parameter values are given
in the appendix.
  The equilibrium values $\underline x ^* $ are given by the equilibrium
condition $\underline f (\underline x ^*, \underline a) =0$, respectively
for limit cycles 
$\underline x^*(t+T) = \underline x^*(t)$
with period $T$. For chaotic attractors the trajectory of the
dynamical system
reaches in the time limit of infinity the attractor
trajectory $\underline x^*(t) $, equally for tori with irrational
winding ratios.
In all cases the stability can be analysed considering small
perturbations $\Delta \underline x (t)$ around the attractor trajectories
\begin{equation}
	\frac{d}{dt} \Delta  \underline x 
        = \left.
        \frac{d \underline f}{d \underline x} 
        \right| _{\underline x  ^*(t)}
        \cdot \Delta \underline x
        \quad.
	\label{dynamicsdeltaf}
\end{equation}
Here, any attractor is notified by $\underline x  ^*(t)$, be it an
equilibrium, periodic orbit or chaotic attractor.
In this ODE system the linearized dynamics is given with the Jacobian
matrix $\frac{d \underline f}{d \underline x} $ of the ODE system
Eq. (\ref{dynamicsf}) evaluated at the trajectory points
$\underline x  ^*(t)$ given in notation of 
$\left.(d \underline f / d \underline x)
\right| _{\underline x  ^*(t)} $.
The Jacobian matrix is analyzed for equilibria
in terms of eigenvalues to determine stability and the loss
of it at bifurcation points, negative real part indicating stability. 
For the stability and loss of it for
limit cylces Floquet multipliers are more common (essentially the 
exponentials of eigenvalues), multipliers inside the unit circle 
indicating stability, and where they leave eventually the unit circle
determining the type of limit cycle bifurcations. 
And for chaotic systems Lyapunov exponents
are determined from the Jacobian around the trajectory, positive 
largest exponents
showing deterministic chaos, zero largest showing limit cycles including tori,
largest smaller zero indicating fixed points.

\subsection{Symmetries}

To investigate the bifurcation structure of the system under investigation
we first observe the symmetries due to the multi-strain structure of the
model. This becomes important for the time being for  
equilibria\footnote{Equilibria are often called fixed points 
in dynamical systems theory,
here we try to avoid this term, since in symmetry  the term \emph{fixed}
is used in a more specific way, see below.}
and limit cycles.
We introduce the following notation:
With a symmetry transformation matrix $\mathbf{S}$
{ \small
\begin{equation}
\mathbf{S}:=\left( 
\begin{array}[c]{c c c c c c c c c c}
    1& 0& 0& 0& 0& 0& 0& 0& 0& 0\\ 
    0& 0& 1& 0& 0& 0& 0& 0& 0& 0\\ 
    0& 1& 0& 0& 0& 0& 0& 0& 0& 0\\ 
    0& 0& 0& 0& 1& 0& 0& 0& 0& 0\\ 
    0& 0& 0& 1& 0& 0& 0& 0& 0& 0\\ 
    0& 0& 0& 0& 0& 0& 1& 0& 0& 0\\ 
    0& 0& 0& 0& 0& 1& 0& 0& 0& 0\\ 
    0& 0& 0& 0& 0& 0& 0& 0& 1& 0\\ 
    0& 0& 0& 0& 0& 0& 0& 1& 0& 0\\ 
    0& 0& 0& 0& 0& 0& 0& 0& 0& 1
\label{symmetrymatrix}
\end{array}
\right)
\end{equation}
}
we have the
following symmetry:
{ \small
\begin{align}\label{eqn:symmatryequi}
\textrm{If  }\underline x ^* =\begin{pmatrix}
  S^*\\
    I_1^*\\
    I_2^*\\
    R_1^*\\
    R_2^*\\
    S_{1}^*\\
    S_{2}^*\\
    I_{12}^*\\
    I_{21}^*\\
  R^*
\end{pmatrix}
\textrm{ is equilibrium or limit cycle, then also  }
\mathbf{S} \underline x ^*=\begin{pmatrix}
  S^*\\
    I_2^*\\
    I_1^*\\
    R_2^*\\
    R_1^*\\
    S_{2}^*\\
    S_{1}^*\\
    I_{21}^*\\
    I_{12}^*\\
  R^*
\end{pmatrix}\;.
\end{align}
}

\noindent
with 
$\underline x^*$ equilibrium values 
or $\underline x^*= \underline x^*(t)$ limit cycle for all 
times $t\in[0,T]$.
For the right hand side $\underline f $ of the ODE system Eq. (\ref{dynamicsf})
the kind of symmetry found above
is called $\mathbb{Z}_2$-symmetry when the following equivariance
condition holds
\begin{align}\label{eqn:equivariancecondition}
  \underline{f}(\mathbf{S}\underline{x},\underline a)
   =\mathbf{S}\underline{f}( \underline{x},\underline a)
\end{align}
with $\mathbf{S}$ a matrix
that obeys $\mathbf{S}\ne \mathbf{I}$ and $\mathbf{S}^2=\mathbf{I}$,
where $\mathbf{I}$ is the unit matrix. Observe that besides
$\mathbf{S}$ also $\mathbf{I}$
satisfies~(\ref{eqn:equivariancecondition}).
The symmetry transformation matrix $\mathbf{S}$ 
in Eq. (\ref{symmetrymatrix}) fulfills these requirements.
It is easy to verify that the $\mathbb{Z}_2$-equivariance
conditions Eq. (\ref{eqn:equivariancecondition}) 
and the properties of $\mathbf{S} $
are satisfied for our ODE system.
In Seydel \cite{Seyd94} a simplified version of the famous Brusselator that
shows this type of symmetry is discussed.  There, an equilibrium and
also a limit cycle show a pitchfork bifurcation with symmetry
breaking.

An equilibrium $\underline x^*$ is called \emph{fixed} 
when $\mathbf{S}\underline x^*=\underline x^*$ (see
\cite{Kuzn2004}).  Two equilibria $\underline x^*,\underline y^*$ 
where $\mathbf{S}\underline x^* \ne \underline x^*$, are
called $\mathbf{S}$-conjugate if their corresponding solutions satisfy
$\underline y^*=\mathbf{S}\underline x^*$ 
(and because $\mathbf{S}^2=\mathbf{I}$ also
$\underline x^*=\mathbf{S}\underline y^*$).
For limit cycles a similar terminology is introduced. A periodic
solution is called \emph{fixed}  when
$\mathbf{S}\underline x^*(t)=\underline x^*(t)$ 
and the associated limit cycles
are also called \emph{fixed} \cite{Kuzn2004}.  
There is another type of periodic
solution that is not fixed but called \emph{symmetric} when

\vspace{-0.8cm}

\begin{align}\label{eqn:defsymmetrylc}
\mathbf{S}\underline x^*(t)=\underline x^* \left( t+\frac{T}{2} \right)
\end{align}
where $T$ is the period. Again the associated limit cycles are also
called \emph{symmetric}. Both types of limit cycles $L$ are
$\mathbf{S}$-invariant as curves : $\mathbf{S}L=L$. That is, in the
phase-plane where time parameterizes the orbit, the cycle and the
transformed cycle are equal. A $\mathbf{S}$-invariant cycle is either
fixed or symmetric.
Two noninvariant limit cycles ($\mathbf{S}L \ne L$) are called
$\mathbf{S}$-conjugate if their corresponding periodic solutions
satisfy $\underline{y}^*(t)=\mathbf{S}\underline{x}^*(t),\; \forall t \in
\mathbb{R}$.
  The properties of the symmetric systems and the introduced terminology
are used below with the interpretation of the numerical bifurcation
analysis results. We refer to \cite{Kuzn2004} for an overview of the
possible bifurcations of equilibria and limit cycles of
$\mathbb{Z}_2$-equivariant systems.


\section{Bifurcation diagrams for various {\boldmath  $\alpha $} values}

We show the results of the bifurcation analysis in bifurcation diagrams
for several $ \alpha  $ values, varying $\phi $ continuously. Besides
the previously investigated case of $ \alpha  = 2 \; year ^{-1}$,
we show also a case of smaller and a case of larger $ \alpha  $ value,
obtaining more information on the bifurcations possible in the model
as a whole. The above mentioned symmetries help in understanding the
present bifurcation structure.

\subsection{Bifurcation diagram for \boldmath {$\alpha = 3$}}

\begin{figure}[htb]
\begin{center}
   a)  \epsfysize=3.3cm
       \epsfbox{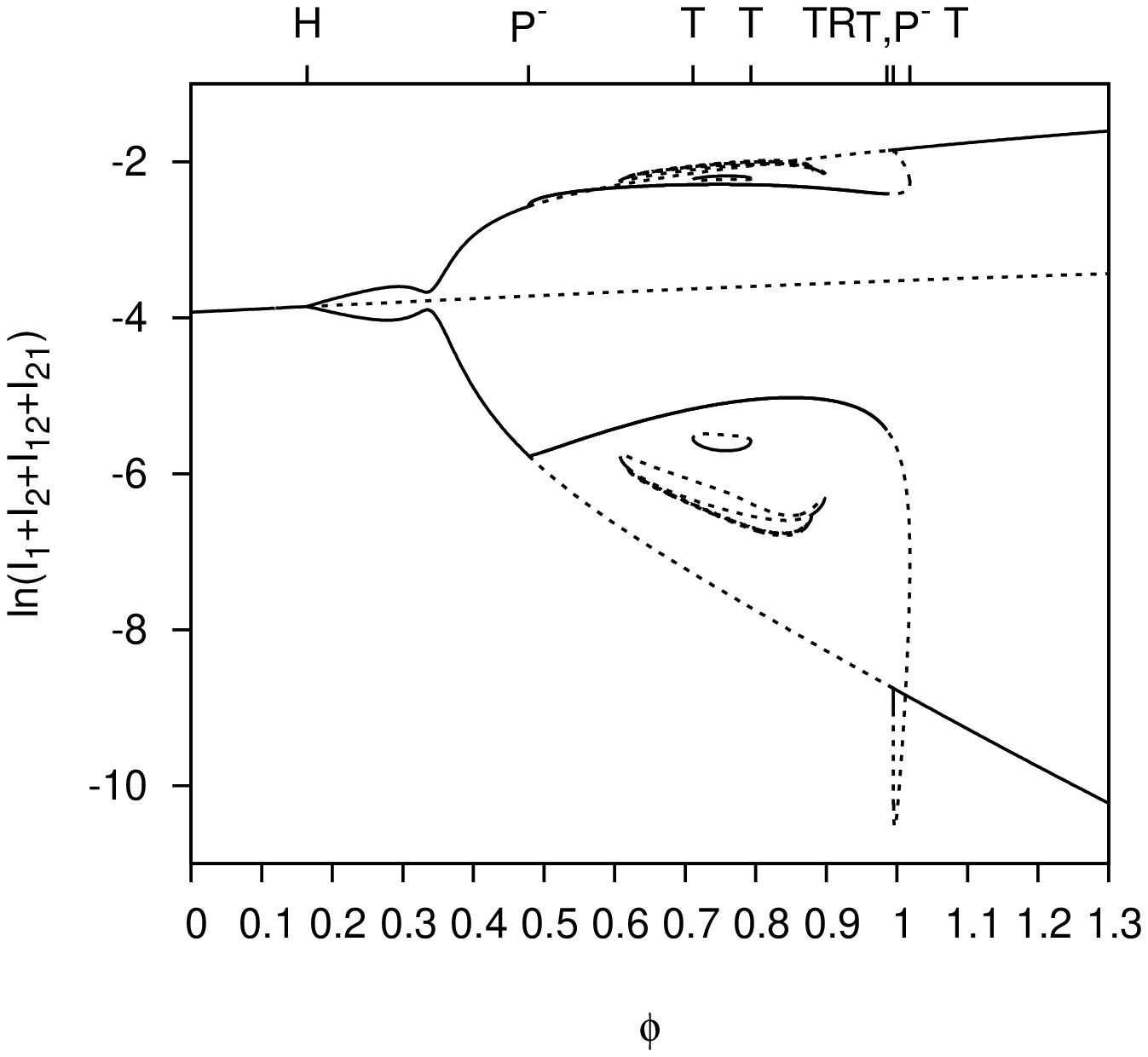} $ \quad  $	
   b)  \epsfysize=3.3cm
       \epsfbox{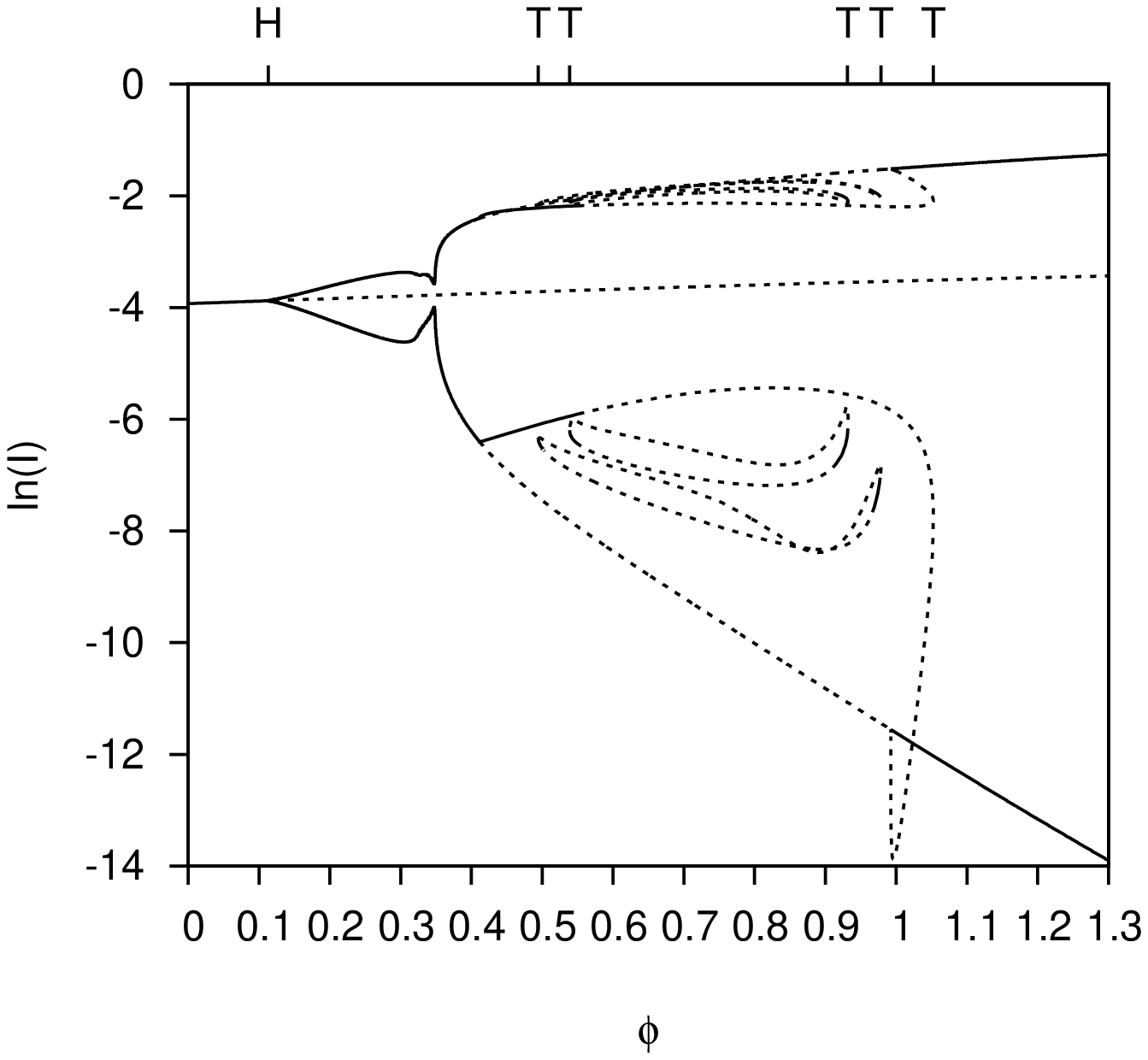} $ \quad  $	
   c)  \epsfysize=3.3cm
       \epsfbox{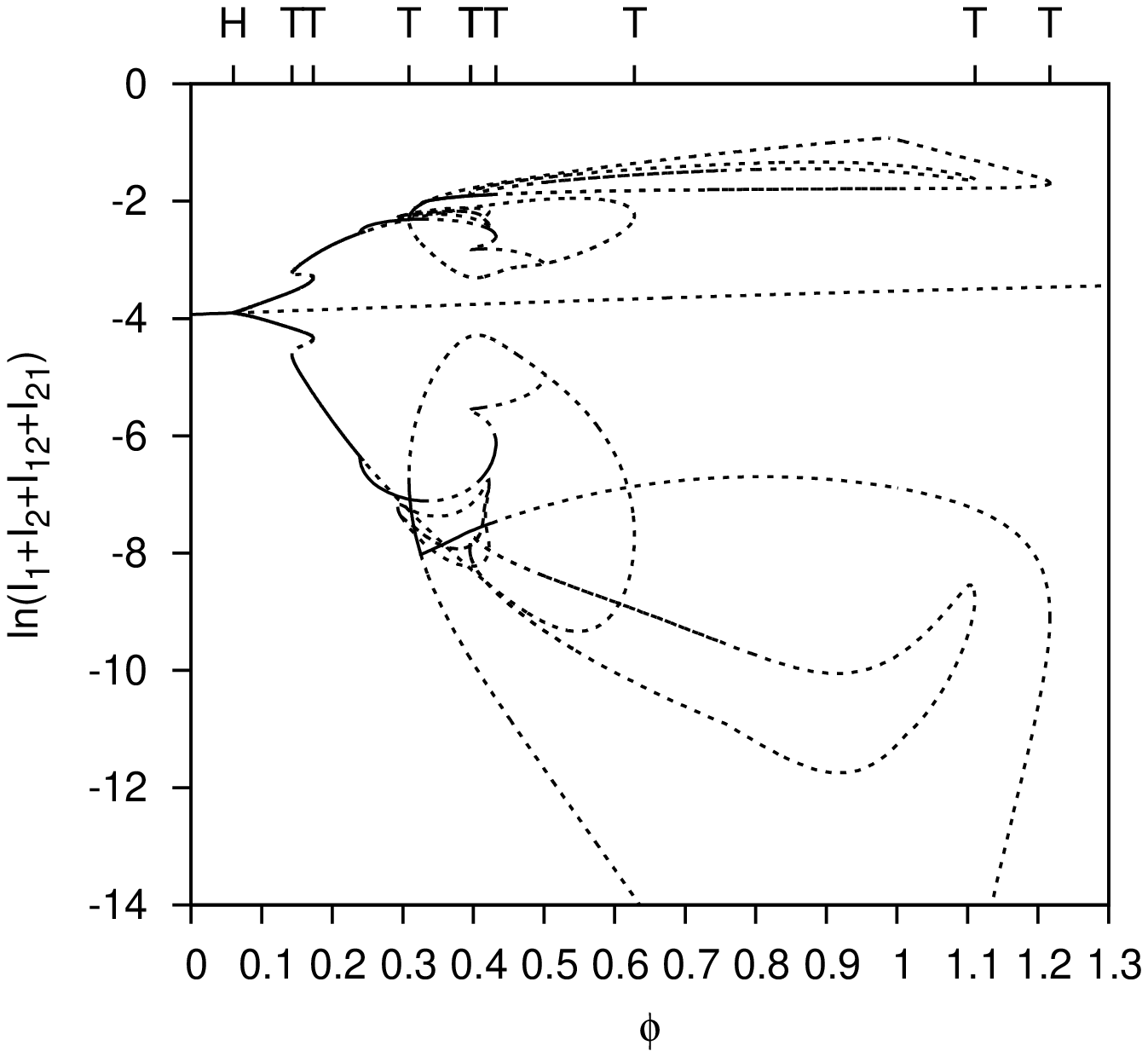} $ \quad  $
\end{center}
\vspace{-0.5cm} 
\caption[]{\label{fig:bifdia1phi} \protect {\small 
    a) 
    $\alpha=3$: Equilibria or extremum values for limit
    cycles for logarithm of
    total infected $I_1+I_2+I_{12}+I_{21}$. Solid lines
    denote stable equilibria or limit cycles, dashed lines unstable
    equilibria or periodic-one limit cycles. Hopf bifurcation $H$ 
    around $\phi=0.16$
    two pitchfork bifurcations $P^- $ and a torus bifurcation $TR $.
    Besides this main bifurcation structure we found coexisting
    tangent bifurcations 
    $T$ 
    between which some of the isolas live, see especially the one 
    between $\phi=0.71$ and $0.79$. Additionally found flip bifurcations
    are not marked here, see text.
    b) 
    $\alpha=2$: In this case we have a Hopf bifurcation
    $H$ at $\phi=0.11$, and besides the similar structure as found in a)
    also more separated tangent bifurcations $T$
    at $\phi=0.494$, $0.539$, $0.931$, $0.978$ and  $1.052$
    c)
    $\alpha=1$: Here we have the Hopf bifurcation at $\phi=0.0598$
    and thereafter many tangent bifurcations $T$, again with
    coexisting limit cylces.
    }}
\end{figure}

For $\alpha=3$ the one-parameter bifurcation diagram is shown in
Fig. \ref{fig:bifdia1phi} a).  
Starting with
$\phi=0$ there is a stable fixed equilibrium,
fixed in the above mentioned notion for symmetric systems. 
This equilibrium becomes
unstable at a Hopf bifurcation $H$ at $\phi=0.164454$.
A stable fixed limit cycle
originates at this Hopf bifurcation. This limit cycle shows a
supercritical pitch-fork bifurcation $P^-$, 
i.e. a bifurcation of
a limit cycle with Floquet multiplier 1, splitting the original limit
cycle into two new ones.
Besides the now unstable
branch two new branches originate for the pair of conjugated limit
cycles. The branches merge again at another supercritical pitch-fork
bifurcation $P^-$,  after which the limit cycle 
is stable again for
higher $\phi$-values.  The pair of $\mathbf{S}$-conjugate limit cycles
become unstable at a torus bifurcation $TR$ at $\phi=0.89539 $.

Besides this main bifurcation pattern we found two isolas, that is an
isolated solution branch of limit cycles \cite{GoSc85}. These isola
cycles $L$ are not $\mathbf{S}$-invariant, that is $\mathbf{S}L\ne L$.
Isolas
consisting of isolated limit cycles exist between two tangent
bifurcations. One isola consists of a stable and an unstable branch.
The other shows more complex bifurcation patterns.  There is no full
stable branch.
For $\phi=0.60809$ at the tangent bifurcation
$T$ a stable and an unstable limit cycle collide. The stable branch
becomes unstable via a flip bifurcation 
or periodic doubling bifurcation $F$, with Floquet multiplier $(-1)$,
at $\phi= 0.61918$ which
is also pitchfork bifurcation for the period-two limit cycles.  At the
other end of that branch at the tangent bifurcation $T$ at
$\phi=0.89768$ both colliding limit cycles are unstable. Close to this
point at one branch there is a torus bifurcation $TR$, also
called Neimark-Sacker bifurcation, at
$\phi=0.89539$ and a flip bifurcation $F$ at $\phi=0.87897$ which is
again a pitchfork bifurcation $P$ for the period-two limit cycles.
Contiuation of the stable branch originating for the flip bifurcation
$F$ at $\phi=0.61918$ gives another flip bifurcation $F$ at
$\phi=0.62070$ and one closed to the other end at $\phi=0.87897$,
namely at $\phi=0.87734$.
These results suggest that for this isola two classical routes to
chaos can exist, namely via the torus or Neimark-Sacker bifurcation where
the dynamics on the originating torus is chaotic, and the cascade of
period doubling route to chaos.

\subsection{Bifurcation diagram for \boldmath {$\alpha = 2$}}

For $\alpha=2$ the one-parameter bifurcation diagram is shown in
Fig. \ref{fig:bifdia1phi} b).
The stable fixed equilibrium
becomes unstable at a supercritical Hopf bifurcation $H$ 
at $\phi=0.1132861$
where a
stable fixed limit cycle originates. This stable limit cycle becomes
unstable at a superciritcal pitchfork bifurcation point $P^-$ 
at $\phi=0.4114478$
for a limit cycle.
This point marks the origin of a pair of $\mathbf{S}$-conjugate stable
limit cycles besides the now unstable fixed limit cycle.
Here one has to consider 
the two infected
subpopulations $I_1$ and $I_2$
to distinguish the conjugate limit cycles. 
Because the two variables $I_1$ and $I_2$ are
interchangeable this can also be interpreted as the stable limit
cycles for the single variable say $I_1$. The fixed stable equilibrium
below the Hopf bifurcation where we have $I_1^*=I_2^*$, $R_1^*=R_2^*$,
$S_1^*=S_2^*$ and $I_{12}^*=I_{21}^*$ is a fixed equilibrium.  For the
fixed limit cycle in the parameter interval between the Hopf
bifurcation and the pitchfork bifurcation we have
${I}_1^*(t)={I}_2^*(t)$, ${R}_1^*(t)={R}_2^*(t)$,
${S}_1^*(t)={S}_2^*(t)$ and
${I}_{12}^*(t)={I}_{21}^*(t)$.  This means that at the Hopf
bifurcation $H$ the stable fixed equilibrium becomes an unstable fixed
equilibrium.
In the parameter interval between the two pitchfork bifurcations
$P^-$ at $\phi=0.4114478$ and subcritical $P^+ $ at $\phi=0.9921416$, two
stable limit cycles coexist and these limit cycles are
$\mathbf{S}$-conjugate.  At the pitchfork bifurcation points the fixed
limit cycle becomes unstable and remains fixed, and two stable
$\mathbf{S}$-conjugate limit cycles originate (see \cite[Theorem
7.7]{Kuzn2004}).
  The invariant plane $I_1=I_2, R_1=R_2, S_1=S_2, I_{12}=I_{21}$ forms
the separatrix between the pair of stable $\mathbf{S}$-conjugate limit
cycles ${x}^*(t)$ and $\mathbf{S}{x}^*(t),\; \forall t \in
\mathbb{R}$.  The initial values of the two state variables $S(t_0)$ and
$R(t_0)$ together with the point on the invariant plane, determine to
which limit cycle the system converges.
Continuation of the stable symmetric limit cycle gives a torus or
Neimark-Sacker bifurcation at point denoted by $TR$
at $\phi= 0.5506880$. At his point the
limit cycles become unstable because a pair of complex-conjugate
multipliers crosses the unit circle.
  Observe that at this point in the time series plot 
\cite[there Fig. 12]{AgSt2007}
the chaotic region
starts. In \cite{AlSp2006} the following route to chaos, namely the
sequence of Neimark-Sacker bifurcations into chaos, is mentioned.
Increasing the bifurcation parameter $\phi$ along the now unstable
pair of $\mathbf{S}$-conjugate limit cycles leads to a tangent
bifurcation $T$ at $\phi=1.052418$
where a pair of two unstable limit cycles collide.
This branch terminates at the second pitchfork bifurcation point
denoted by $P^+$ at $\phi=0.9921416$. 
Because the first fold point gave rise to a stable
limit cycle and this fold point to an unstable limit cycle we call the
first pitchfork bifurcation supercritical and the latter pitchfork
bifurcation subcritical.
These results agree very well with the simulation results shown in the
bifurcation diagram for the maxima and minima of the overall infected
\cite[there Fig. 15]{AgSt2007}. Notice that AUTO \cite{autoreference}
calculates only the global extrema during a cycle, not the local
extrema.
Fig.  \ref{fig:bifdia1phi} b) shows also two isolas
similar to those for $\alpha = 3$ in Fig. \ref{fig:bifdia1phi}
a).

\subsection{Bifurcation diagram for \boldmath {$\alpha = 1$}}

For $\alpha=1$ the bifurcation diagram is shown in
Fig \ref{fig:bifdia1phi} c). 
In the lower $\phi$ parameter range there
is bistability of two limit cycles in an interval bounded by two
tangent bifurcations $T$. The stable manifold of the intermediate
saddle limit cycle acts as a separatrix.  Inceasing $\phi$ the stable
limit cycles become unstable at the pitchfork bifurcation $P$ at
$\phi=0.2390695$.
  Following the unstable primary branch, for larger values of $\phi$ we
observe an open loop bounded by two tangent bifurcations $T$. The
extreme value for $\phi$ is at $\phi=0.6279042$.  Then lowering $\phi$
there is a pitchfork bifurcation $P$ at $\phi=0.5016112$. Later we
will return to the description of this point.  Lowering $\phi$ further
the limit cycle becomes stable again at the tangent bifurcations $T$ at
$\phi=0.3086299$.  Increasing $\phi$ this limit cycle becomes unstable
again at the pitchfork bifurcation $P$ at $\phi=0.3253242$.

Continuation of the secondary branch of the two
$\mathbf{S}$-conjugated limit cycles from this point reveals that the
stable limit cycle becomes unstable at a torus bifurcation $TR$ at
$\phi=0.4257346$.  
The simulation results
depicted in \cite[Fig. 13]{AgSt2007} show that there is chaos beyond
this point.
  The secondary pair of $\mathbf{S}$-conjugate limit cycles that
originate from pitchfork bifurcation $P$ at $\phi=0.2390695$ becomes
unstable at a flip 
bifurcation $F$.  Increasing
$\phi$ further it becomes stable again at a flip bifurcation $F$.
Below we return to the interval between these two flip bifurcations.
The stable part becomes unstable at a tangent bifurcation $T$, then
continuing, after a tangent bifurcation $T$ and a Neimark-Sacker
bifurcation $TR$. This bifurcation can lead to a sequence of
Neimark-Sacker bifurcations into chaos. The unstable limit cycles
terminates via a tangent bifurcation $F$ where the primary limit cycle
possesses a pitchfork bifurcation $P$ at $\phi=0.5016112$. 
  At the flip bifurcation $F$ the cycle becomes unstable and a new
stable limit cycle with double period emanates. The stable branch
becomes unstable at a flip bifurcation again. 
We conclude that there is a cascade of
period doubling route to chaos. Similarly this happens in reversed
order ending at the flip bifurcation where the secondary branch
becomes stable again.

\begin{figure} 
\begin{center}
   a)  \epsfysize=3.3cm
       \epsfbox{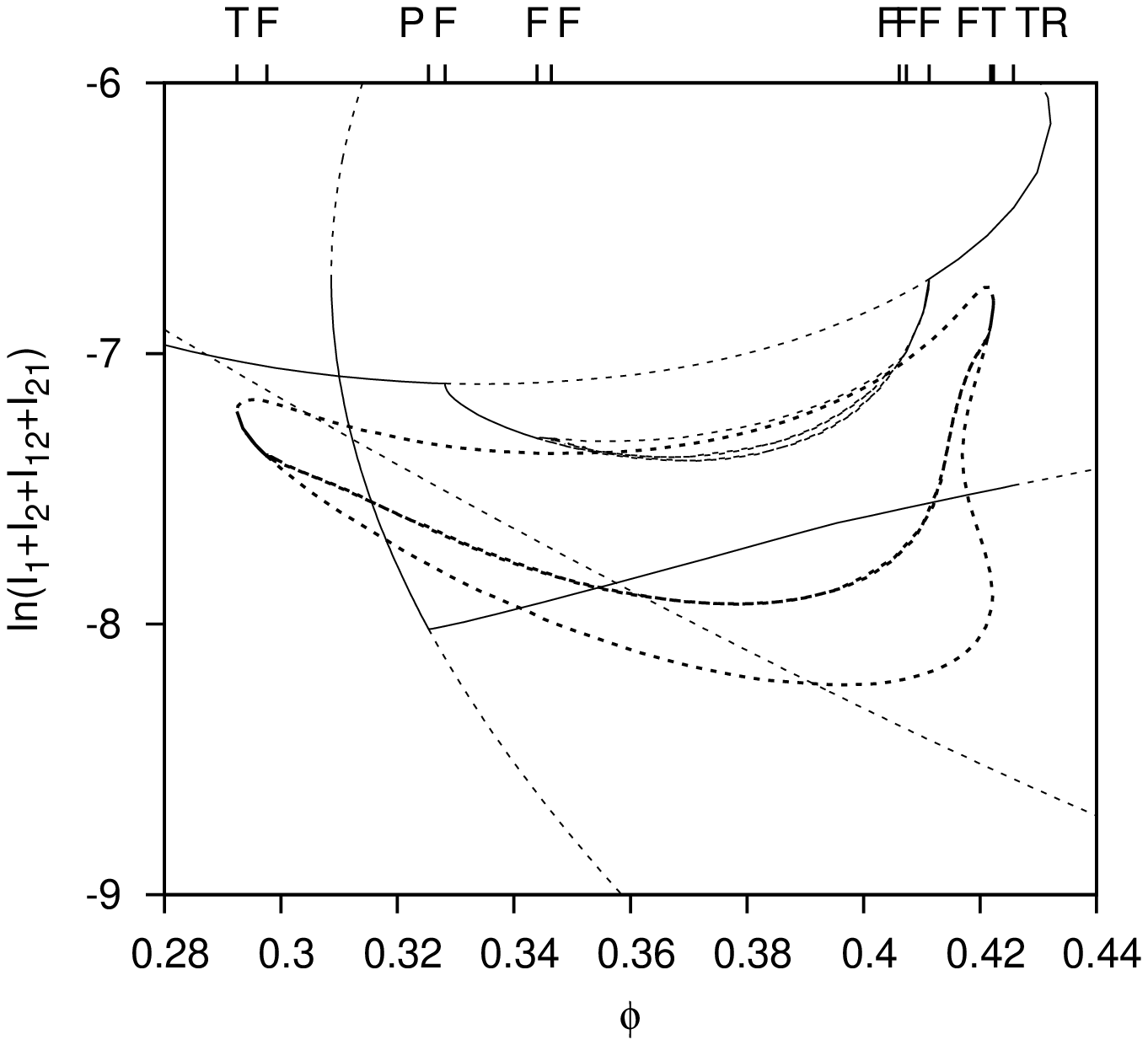} $ \quad  $	
   b)  \epsfysize=3.3cm
       \epsfbox{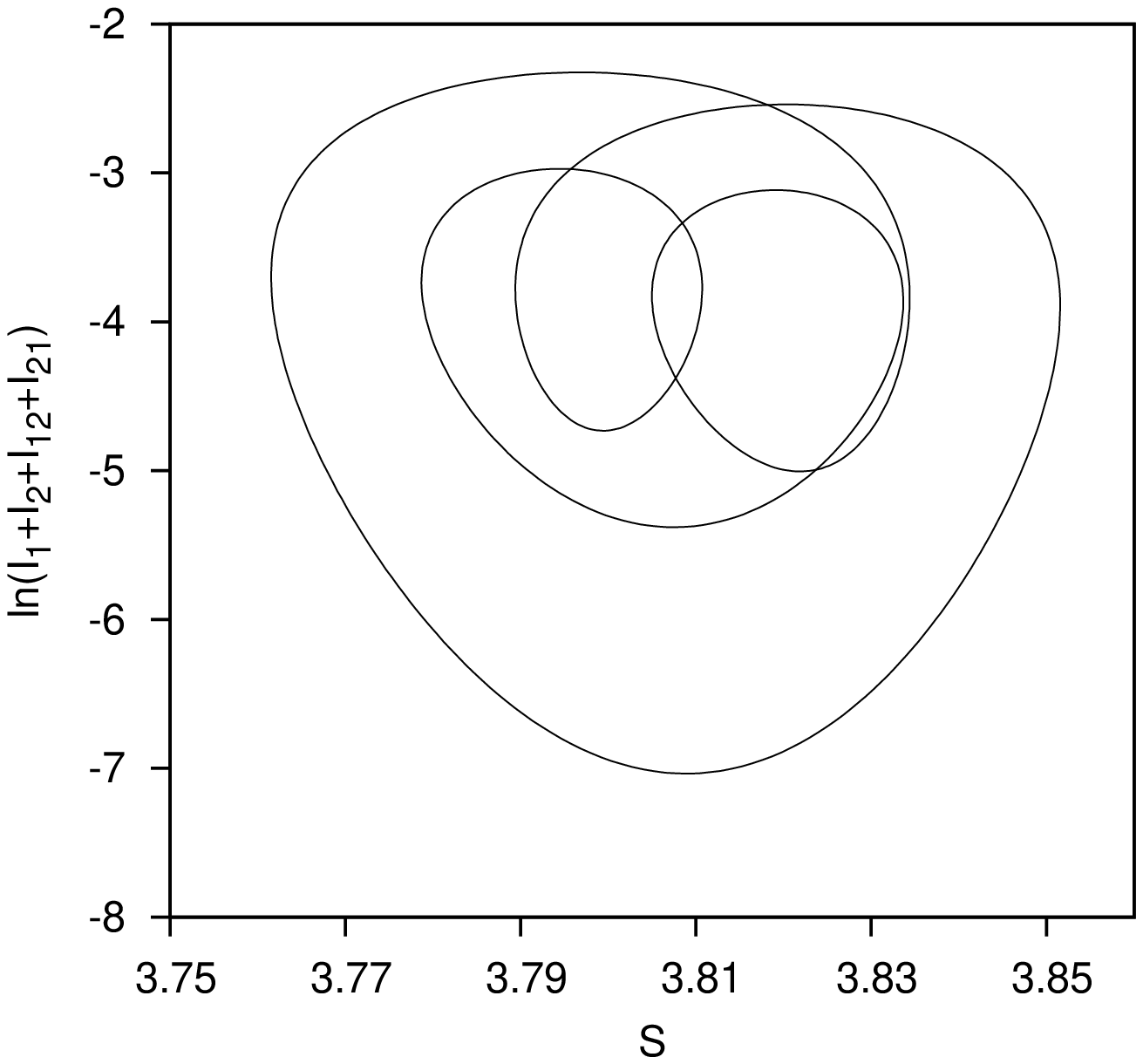} $ \quad  $	
   c)  \epsfysize=3.3cm
       \epsfbox{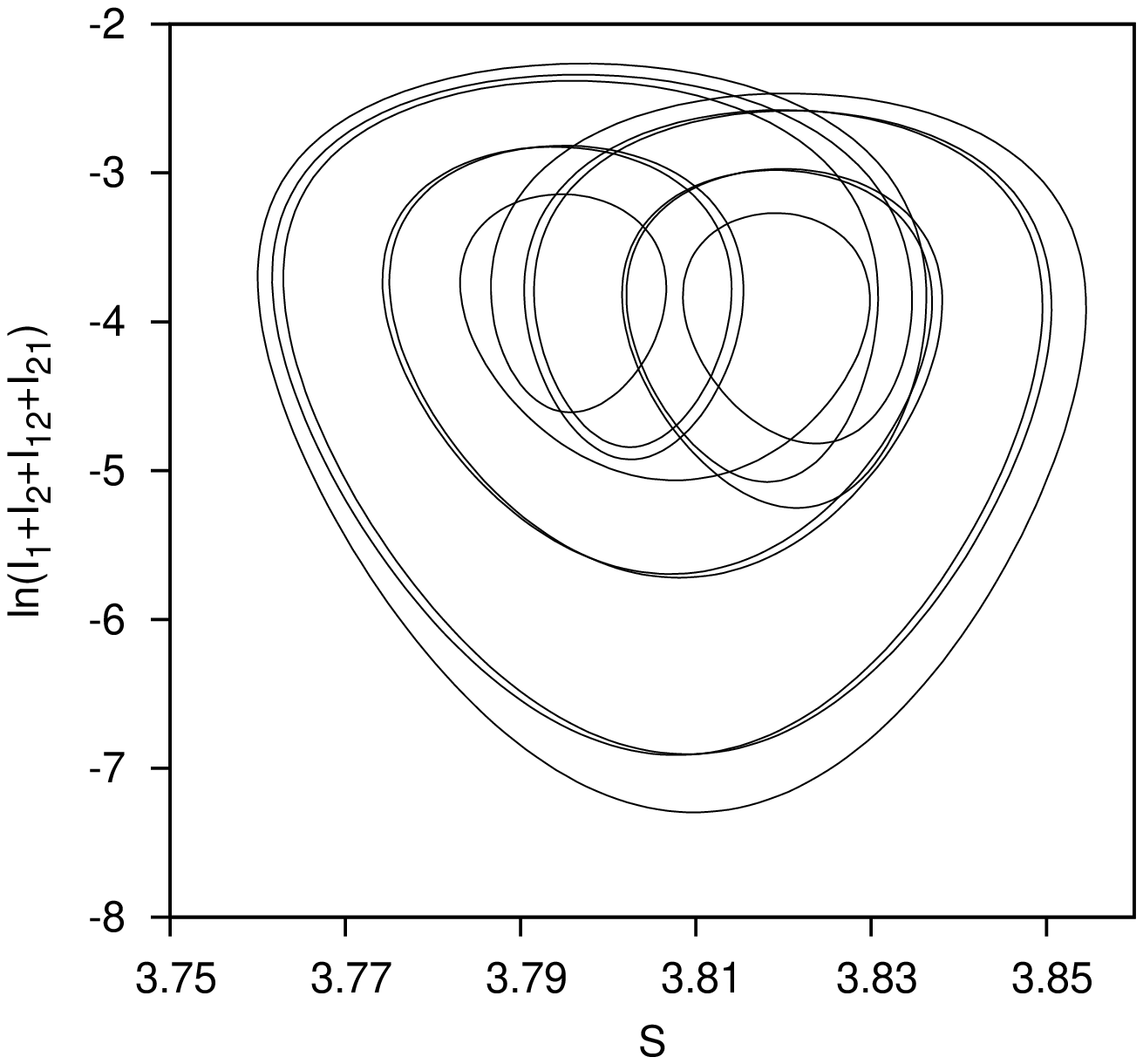} $ \quad  $
\end{center}
\vspace{-0.5cm} 
\caption[]{\label{fig:bifdia1iphidd} \protect {\small 
    a) $\alpha=1$. Detail of Fig. \ref{fig:bifdia1phi} c). We find  
    pitchfork bifurcations $P$ at $\phi=0.239$ and $0.325$, flip
    bifurcations $F$ at $\phi=0.298$, $0.328$,$0.344$,$0.346$,
    $0.406$, $0.407$, $0.411$ and  $0.422$, further tangent bifurcations
    $T$ at $\phi=0.292$, $0.346$ and  $0.422$.
    Four almost coexisting bifurcations, namely $F$'s
    at $\phi=0.4112590$.
    b) and c) 
    state space-plots of susceptibles and 
    logarithm of
    infected 
    for $\alpha=1$ and $\phi=0.294$ for two coexisting
    stable limit cycles.
  }}
\end{figure}

Fig. \ref{fig:bifdia1iphidd} a) gives the results for the interval
$0.28\le \phi \le 0.44$ where only the minima are show. In this plot
also a ``period three'' limit cycle is shown. In a small region it is
stable and coexists together with the ``period one'' limit cycle. 
The
cycles are shown in Fig. \ref{fig:bifdia1iphidd} b) and c) for
$\phi=0.294$.   
The one in c) looks like a period-3 limit cycle. 
In Fig. \ref{fig:bifdia1iphidd} continuation of the
limit cycle gives a closed graph bounded at the two ends by trangent
bifurcations $T$ where a stable and an unstable limit cycle collide.
The intervals where the limit cycle is stable, are on the other end
bounded by flip bifurcations $F$.  One unstable part intersects the
higher period cycles that originate via the cascade of period doubling
between the period-1 limit cycle flip bifurcations $F$ at
$\phi=0.3281636$ and $\phi=0.4112590$. This suggest that the period-3
limit cycle is associated with a ``period-3 window'' of the chaotic
attractor.  We conjecture that this interval is bounded by two
homoclinic bifurcations for a period-3 limit cycle (see
\cite{BoKoKo98,BoKoKo99,KoBo2002,KoKuKo2004}). 
The bifurcation diagram
shown in \cite[there Fig. 13]{AgSt2007} shows the point where the
chaotic attractor disappears abruptly, possible at one of the two
homoclinic bifurcations. In that region the two conjugated limit
cycles that originate at the pitchfork bifurcation $P$ at
$\phi=0.3253242$ are the attractors.
These results suggest that there are chaotic attractors associated
with the period-1 limit cycle, one occurs via a cascade of flip
bifurcations originating from the two ends at $\phi=0.3281636$ and
$\phi=0.4112590$ and one via a Neimark-Sacker bifurcation $TR$ at
$\phi=0.4257346$.


\section{Two-parameter diagram}

\begin{figure}[htb] 
\begin{center}
\epsfysize=4.0cm
       \epsfbox{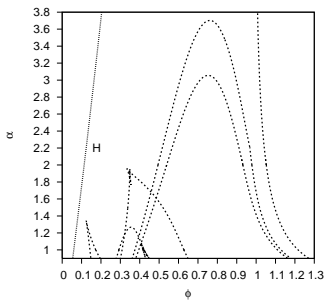}
\end{center}
\vspace{-0.7cm} 
\caption[]{\label{fig:phialphad} \protect {\small 
    Two-dimensional parameter bifurcation diagram with $\phi$ and
    $\alpha$ as parameters. Only one Hopf bifurcation (dotted lines)
    and many tangent bifurcation curves (dashed lines) are shown in the range
    $\alpha\in[1,4]$. The isolated limit cycles originate above
    $\alpha=3$. For lower values of $\alpha$ periodic doubling routes
    to chaos originate.}}
\end{figure}

We will now link the three studies of the different $\alpha $ values
by investigating a two-parameter diagram for $\phi$ and $\alpha $,
concentrating especially on the creation of isolated limit cycles,
which sometimes lead to further bifurcations inside the isola region.
  Fig. \ref{fig:phialphad} gives a two-parameter bifurcation diagram
where $\phi$ and $\alpha$ are the free parameters. 
For low
$\phi$-values there is the Hopf bifurcation $H$ and all other curves
are tangent bifurcation curves.

Isolas  appear or disappears upon crossing an
isola variety. At an elliptic isola point an isolated solution branch
is born, while at a hyperbolic
isola point an isolated solution branch vanishes by coalescence with
another branch \cite{GoSc85}. 
From Fig.~\ref{fig:phialphad} we see that 
at two values of $\alpha>3$
isolas are born. 
Furthermore, period doubling bifurcations
appear for lower $\alpha  $ values, indicating the Feigenbaum route 
to chaos. However, only the calculation of Lyapunov exponents, which are
discussed
in the next section, can clearly
indicate chaos.


\section{Lyapunov spectra for various {\boldmath  $\alpha $} values}

The Lyapunov exponents are the logarithms of the eigenvalues
of the Jacobian matrix along the
integrated trajectories,  Eq. (\ref{dynamicsdeltaf}), in the limit of
large integration times. Besides for very simple iterated maps no
analytic expressions for chaotic systems can be given for the
Lyapunov exponents.
For the calculation of the iterated Jacobian matrix and its eigenvalues, we
use the QR decomposition algorithm \cite{Eckmannetal1986}.

\epsfigthree{3.3cm}{3.3cm}{3.3cm}{htb}{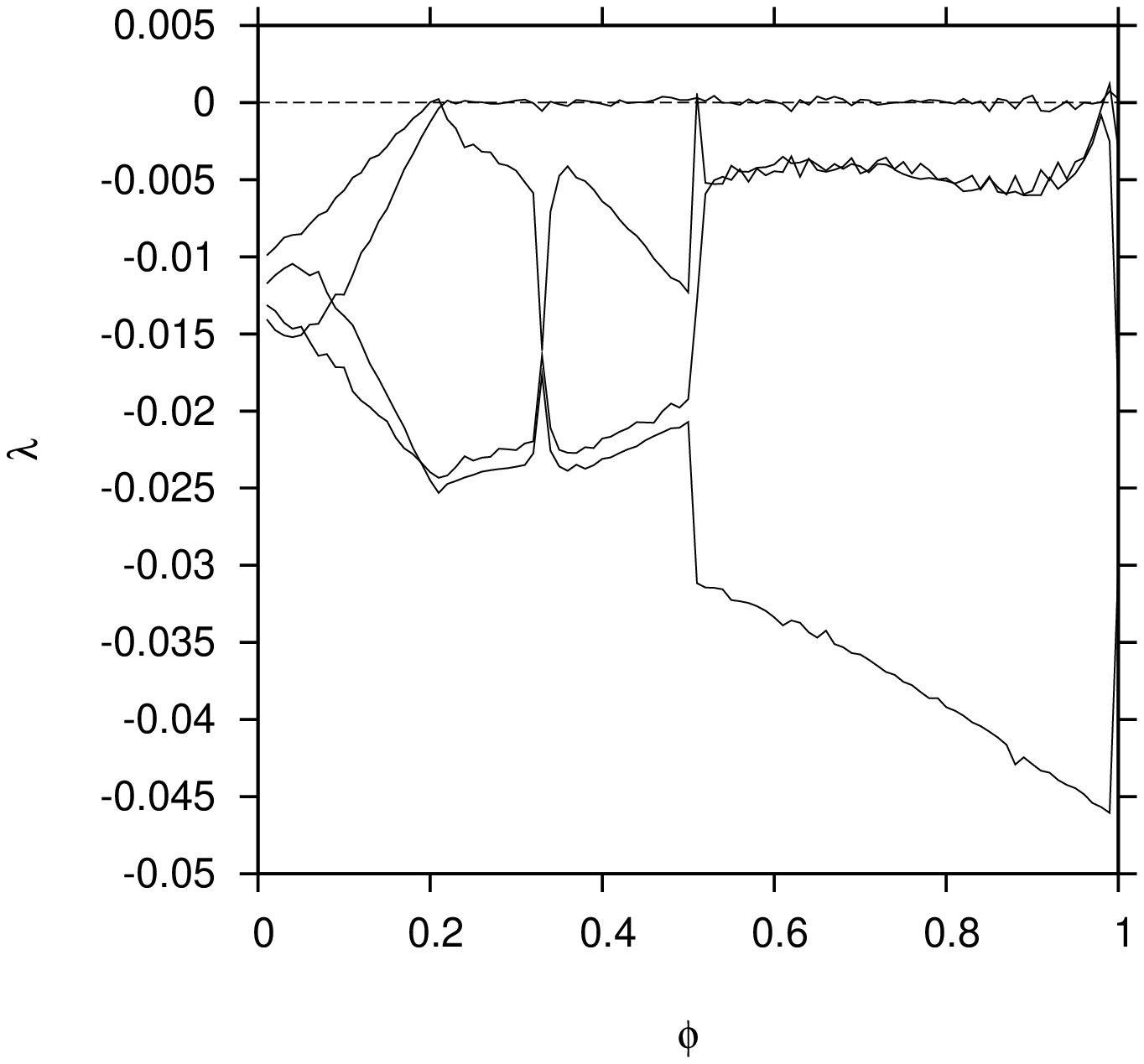}{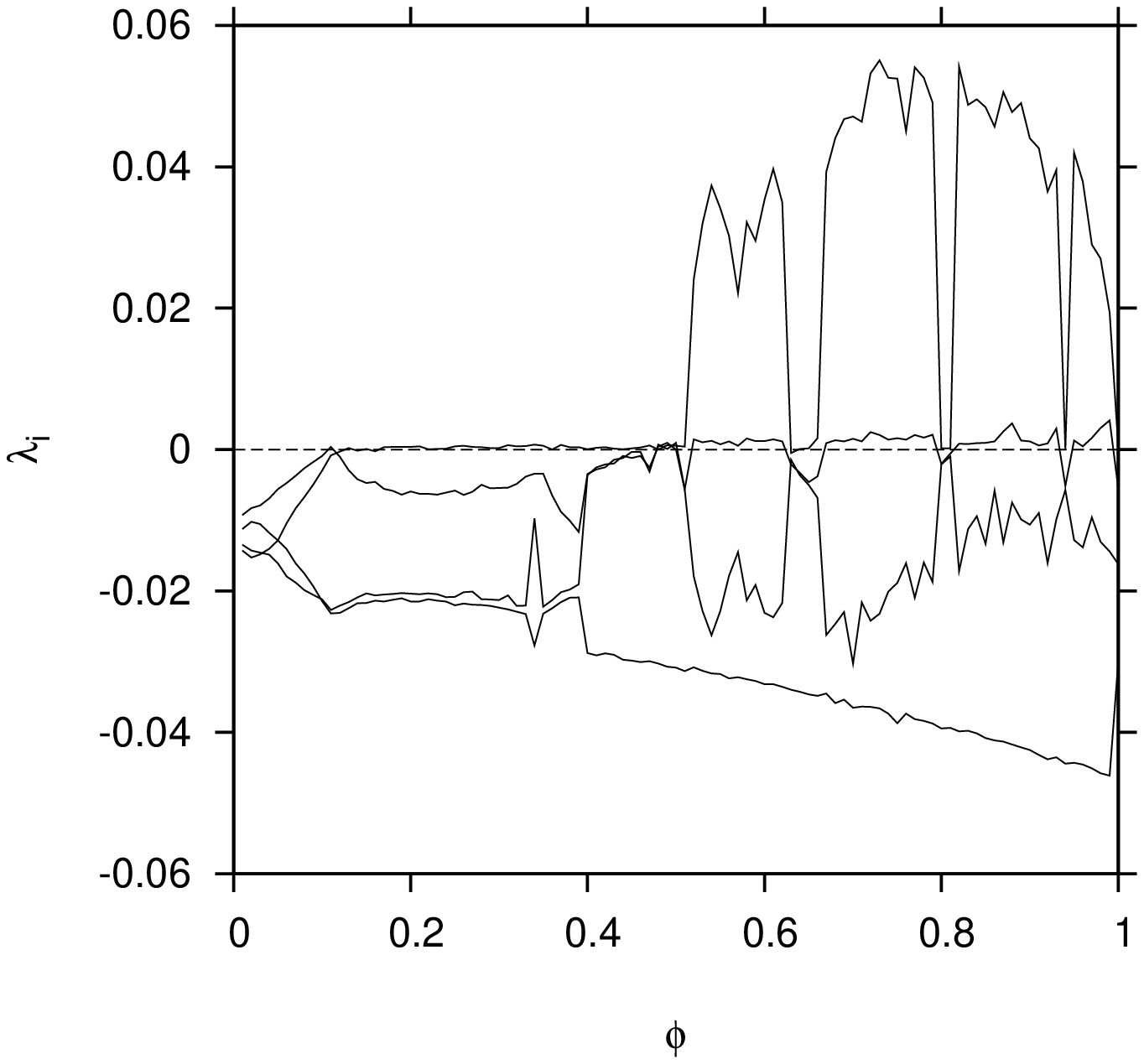}{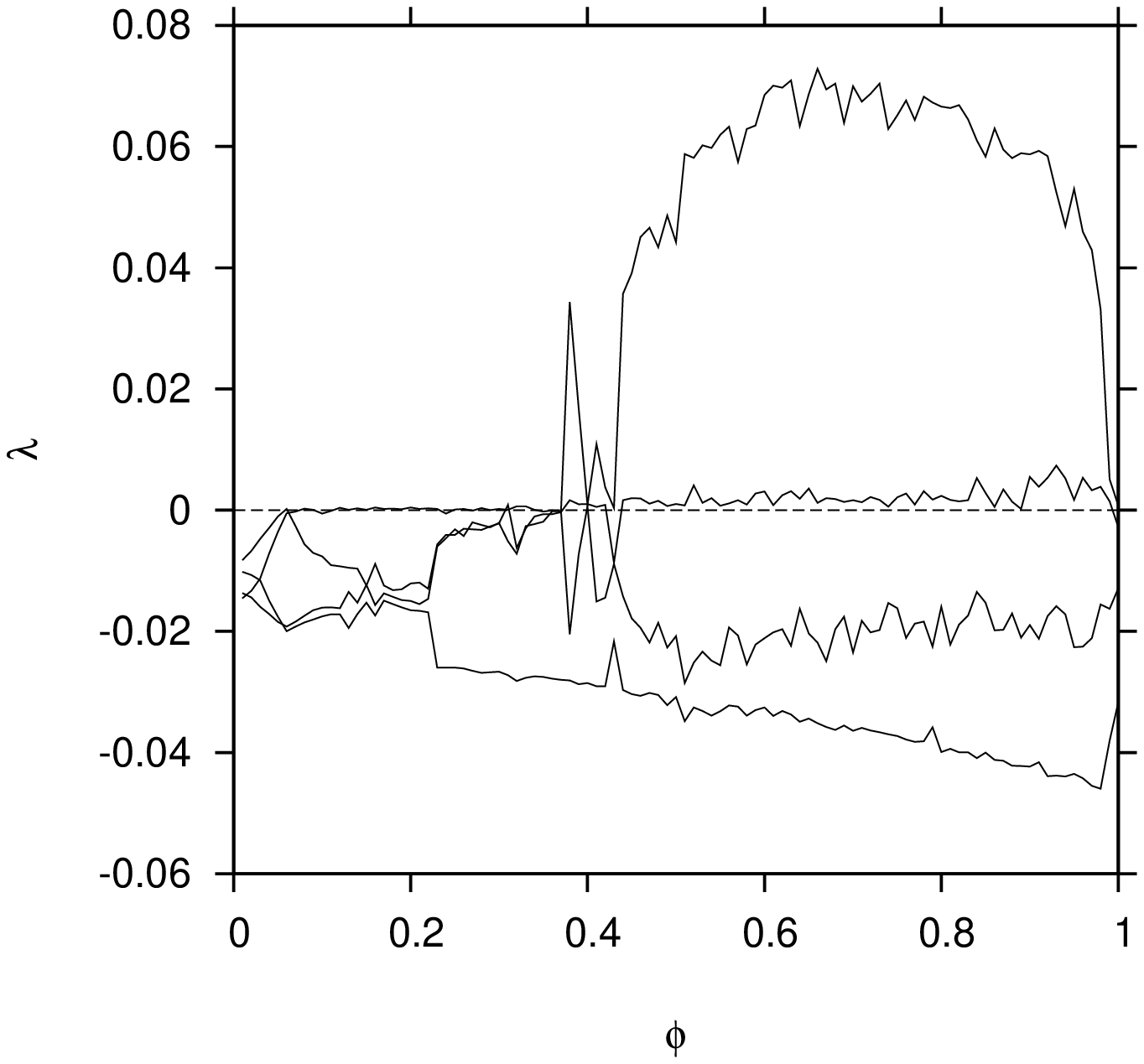}
	{\small Spectrum of the four largest Lyapunov exponents
          with changing parameter $\phi $ and a) fixed $\alpha =4 $,
          b) $\alpha =2 $ and c) $\alpha =1 $.
	}

In Fig. \ref{fig:lyapspect6} we show for various $\alpha $ values
the four largest Lyapunov exponents in the $\phi $ range between zero and
one. For $\alpha =4  $ in Fig. \ref{fig:lyapspect6} a) we see
for small $\phi $ values fixed point behaviour indicated by a negative
largest Lyapunov exponent up to around $\phi=0.2 $. There,
at the Hopf bifurcation point, the largest Lyapunov
exponent becomes zero, indicating limit cycle behaviour for the whole range
of $\phi $, apart from the final bit before $\phi=1 $, where a small spike
with positive Lyapunov exponent might be present, but difficult to
distinguish from the noisy numerical background.

For $\alpha =2 $ in Fig. \ref{fig:lyapspect6} b) however, we see a large
window with positive largest Lyapunov exponent, well separated from the
second largest being zero. This is s clear sign of deterministically
chaotic attractors present for this $\phi $ range. Just a few
windows with periodic attractors, indicated by the zero largest Lyapunov
exponent are visible in the region of $0.5 <\phi <1 $. For smaller 
$\phi $ values we observe qualitatively the same behaviour as already
seen for $\alpha =4  $.
For the smaller value of $\alpha =1 $ in Fig. \ref{fig:lyapspect6} c)
the chaotic window is even larger than for $\alpha =2 $. Hence deterministic
chaos is present for temporary cross immunity in the range around
$\alpha =2 \; year^{-1}$ in the range of $\phi $ between zero and one.

\section{Conclusions}

We have presented a detailed bifurcation analysis for a multi-strain
dengue fever model in terms of the ADE parameter $\phi $, 
in the previously not well investigated region between zero and one,
and a parameter
for the temporary cross immunity $\alpha $.
The symmetries implied by the strain structure, are
taken into account in the analysis.
Many of the possible bifurcations
of equilibria and limit cycles of $\mathbb{Z}_2$-equivariant systems
can be distinguished.
Using AUTO
\cite{autoreference} the different dynamical structures were
calculated.
Future 
time series analysis of epidemiological data has good chances to
give insight into the relevant parameter values purely on topological
information of the dynamics, rather than classical parameter estimation
of which application is in general restricted 
to farely simple dynamical scenarios.

\section*{Acknowledgements}

This work has been supported by the European Union under the Marie Curie
grant MEXT-CT-2004-14338. We thank Gabriela Gomes and Luis Sanchez, 
Lisbon, for scientific support.

%
%



\section{Appendix: Epidemic model equations}

The complete system of ordinary differential equations 
for a two strain epidemiological system allowing for differences in
primary versus secondary infection and temporary cross immunity
is given by 
{ \small
\begin{eqnarray}
    \frac{d}{dt}S& = &   - \frac{\beta}{N} S (I_1+ \phi I_{21}) - \frac{\beta}{N} S (I_2 + \phi I_{12}) + \mu (N-S) \nonumber
\\\nonumber
	\frac{d}{dt}I_1& = &   \frac{\beta}{N}S (I_1+ \phi I_{21}) - (\gamma + \mu)I_1 \nonumber
\\\nonumber
	\frac{d}{dt}I_2& = &   \frac{\beta}{N}S (I_2+ \phi I_{12}) - (\gamma + \mu)I_2 \nonumber
\\\nonumber
	\frac{d}{dt}R_1& = &   \gamma I_1 - (\alpha +\mu) R_1  \nonumber
\\
	\frac{d}{dt}R_2& = &   \gamma I_2 - (\alpha + \mu ) R_2 
        \label{ODE2strain}
\\ \nonumber
	\frac{d}{dt}S_1& = &  - \frac{\beta}{N} S_1 (I_2 + \phi I_{12}) + \alpha R_1 - \mu S_1 \nonumber
\\\nonumber
	\frac{d}{dt}S_2& = &   - \frac{\beta}{N} S_2 (I_1 + \phi I_{21}) + \alpha R_2 - \mu S_2\nonumber
\\\nonumber
	\frac{d}{dt}I_{12}& = &   \frac{\beta}{N}S_1 (I_2+ \phi I_{12}) - (\gamma + \mu)I_{12}\nonumber
\\\nonumber
	\frac{d}{dt}I_{21}& = &   \frac{\beta}{N}S_2 (I_1+ \phi I_{21}) - (\gamma + \mu)I_{21}\nonumber
\\\nonumber
	\frac{d}{dt}R& = &   \gamma (I_{12} +  I_{21}) - \mu R \nonumber
\quad.
\end{eqnarray}
}
For two different strains, $1$ and $2$, we label the SIR classes 
for the hosts that have seen the individual strains.
Susceptibles to both strains (S) get infected with strain $1$ ($I_1$) 
or strain $2$ ($I_2$), with infection rate $\beta$.   
They recover from infection with strain $1$ 
(becoming temporary cross-immune $R_1$) 
or from strain $2$ (becoming $R_2$), with recovery rate $\gamma$ etc.. 
With rate $\alpha$, the $R_1$ and $R_2$ enter again in the 
susceptible classes ($S_1$ being immune against strain 1 but susceptible to 2,
respectively $S_2$), where the index 
represents the first infection strain. 
  Now, $S_1$ can be reinfected 
with strain $2$ (becoming $I_{12}$), 
meeting $I_2$ with infection rate $\beta$ 
or meeting $I_{12}$ with infection rate $\phi\beta$, 
secondary infected contributing differently to the force of infection
than primary infected, etc..

  We include demography of the host population denoting the birth 
and death rate  by $\mu $. 
For constant population size $N$ we have
for the immune to all strains
$R=N-(S+I_1+I_2+R_1+R_2+S_1+S_2+I_{12}+I_{21})$ and therefore we only
need to consider the first 9 equations of Eq. 
(\ref{ODE2strain}), giving 9 Lyapunov exponents.
  In our numerical studies we take the population
size equal to $N=100$ so that numbers of susceptibles, infected etc.
are given in percentage.
As fixed parameter values we take $\mu=(1/65) \; year^{-1} $, 
$\gamma =52 \; year^{-1}$, $\beta = 2\cdot \gamma $. The parameters
$\phi $ and $\alpha $ are varied.

\end{document}